\documentclass[10pt]{article}
\pdfoutput=1
\usepackage{amsthm,graphicx}
\usepackage[letterpaper,margin=1in]{geometry}
\usepackage{xcolor,authblk}

\usepackage{moreverb,url,float,booktabs}

\usepackage[colorlinks,bookmarksopen,bookmarksnumbered,linkcolor=blue,citecolor=blue,urlcolor=blue,breaklinks=true]{hyperref}
\usepackage{breakcites}

\newcommand\BibTeX{{\rmfamily B\kern-.05em \textsc{i\kern-.025em b}\kern-.08em
T\kern-.1667em\lower.7ex\hbox{E}\kern-.125emX}}

\def\volumeyear{2016}
\newcommand\xingfu[1]{{\color{black}#1}}  
\newcommand{\keywords}[1]{Keywords: \textit{#1}}

\begin{document}
\title{Integrating ytopt and libEnsemble to Autotune OpenMC}
\author[1]{Xingfu Wu}
\author[1]{John R. Tramm}
\author[1]{Jeffrey Larson}
\author[1]{John-Luke Navarro}
\author[2]{Prasanna Balaprakash}
\author[1]{Brice Videau}
\author[1]{Michael Kruse}
\author[1]{Paul Hovland}
\author[1]{Valerie Taylor}
\author[3]{and Mary Hall}
\affil[1]{Argonne National Laboratory, Lemont, IL 60439}
\affil[2]{Oak Ridge National Laboratory, Oak Ridge, TN  37830}
\affil[3]{University of Utah, Salt Lake City, UT 84103}
\affil[ ]{~\linebreak\texttt{xingfu.wu@anl.gov}}
\maketitle
\begin{abstract}
ytopt is a Python machine-learning-based autotuning software package developed within the ECP 
PROTEAS-TUNE project. The ytopt software adopts an asynchronous search framework that consists of sampling a small number of input parameter configurations and progressively fitting a surrogate model over the input-output space until exhausting the user-defined maximum number of evaluations or the wall-clock time.  libEnsemble is a Python toolkit for coordinating workflows of asynchronous and dynamic ensembles of calculations across massively parallel resources developed within the ECP PETSc/TAO project. libEnsemble helps users take advantage of massively parallel resources to solve design, decision, and inference problems and expands the class of problems that can benefit from increased parallelism. In this paper we present our methodology and framework to integrate ytopt and libEnsemble to take advantage of massively parallel resources to accelerate the autotuning process. Specifically, we focus on using the proposed framework to autotune the ECP ExaSMR application OpenMC, an open source Monte Carlo particle transport code. OpenMC has seven tunable parameters  some of  which have large ranges such as the number of particles in-flight, which is in the range of 100,000 to 8 million, with its default setting of 1 million. Setting the proper combination of these parameter values to achieve the best performance is extremely time-consuming. Therefore, we apply the proposed framework to autotune the MPI/OpenMP offload version of OpenMC based on a user-defined metric such as the figure of merit (FoM) (particles/s) or energy efficiency energy-delay product (EDP) on Crusher at Oak Ridge Leadership Computing Facility. The experimental results show that we achieve  improvement up to 29.49\% in FoM and up to 30.44\% in EDP.
\end{abstract}
\twocolumn
\section{Introduction}
As we have entered the exascale computing era, high performance, power, and energy management are still key design points and constraints for any next generation of large-scale high-performance computing (HPC) systems. The efficient utilization of power and the optimization of scientific applications under power and energy constraints poses significant challenges. These challenges arise from dynamic phase behavior, manufacturing variation, and increasing system-level heterogeneity. With the growing complexity of HPC ecosystems (hardware stack, software stack, applications), achieving optimal performance and energy efficiency becomes increasingly demanding. The number of tunable parameters that HPC users can configure at the system and application levels has increased significantly, resulting in a dramatically increased parameter space. Exhaustively evaluating all parameter combinations becomes very time-consuming. 
Therefore, autotuning for automatic exploration of the parameter space is necessary.

Autotuning is an approach that explores a search space of tunable parameter configurations of an application efficiently executed on an HPC system. Typically, one selects and evaluates a subset of the configurations on the target system and/or uses analytical models to identify the best implementation or configuration for performance or energy within a given computational budget. 
Such methods are becoming too difficult in practice, however, because of the complexity of hardware, software, and application. Recently, the use of advanced search methods that adopt mathematical optimization methods to explore the search space in an intelligent way has received significant attention in the autotuning community. Such a strategy, however, requires search methods to efficiently navigate the large parameter search space of possible configurations in order to avoid a large number of time-consuming application executions to determine high-performance configurations. 

In the 2022 Exascale Computing Project (ECP) \cite{R46}) annual meeting panel ``Exascale Computing: Hits and Misses,'' Jack Dongarra from the University of Tennessee,
Knoxville, mentioned that one miss is autotuning. As predicted, autotuning should be ready for exascale systems, but  considerable work is still needed. 
Traditional autotuning methods are built on heuristics that derive from automatically tuned BLAS libraries \cite{R23}, experience \cite{R24, R25, R26}, and model-based methods \cite{R27, R28, R29, R30}. At the compiler level \cite{R31}, methods based on machine-learning (ML) are used for automatic tuning of the iterative compilation process \cite{R32} and tuning of compiler-generated code \cite{R33, R34}. Autotuning OpenMP codes went beyond loop schedules to look at parallel tasks and function inlining \cite{R35, R36, R37, R38}. Recent work on leveraging Bayesian optimization to explore the parameter space search showed the potential for autotuning on CPU systems \cite{R39, R2, R40, R41} and on GPU systems \cite{R42, R43}.  Most of the autotuning frameworks mentioned above were for autotuning on a single node or a few compute nodes using only the application runtime as the performance metric. Recently, new autotuning frameworks are emerging for multinode autotuning.  For example, GPtune \cite{R40} autotuned some MPI applications on up to 64 nodes with 2,048 cores with multitask learning using MPI, and Bayesian optimization was applied to increase the energy efficiency of a GPU cluster system \cite{R42}. 

Our recent effort, ytopt \cite{R1, R2, R3}, which is one of the ECP PROTEAS-TUNE projects \cite{R44}, is an ML-based autotuning software package that consists of sampling a small number of input parameter configurations, evaluating them, and progressively fitting a surrogate model over the input-output space until exhausting the user-defined maximum number of evaluations or the wall-clock time. ytopt had already been applied to autotune PolyBench benchmarks with LLVM Clang/Polly loop optimization pragmas and hyperparameter optimization for deep learning applications \cite{R2} and Apache TVM-based scientific applications \cite{R4}. In our previous work \cite{R1} we used ytopt to autotune four hybrid MPI/OpenMP ECP proxy applications \cite{R21} at large scales \xingfu{on large-scale HPC systems Cray XC40 Theta \cite{R49} at Argonne National Laboratory and the IBM Power9 heterogeneous system Summit \cite{R50} at Oak Ridge National Laboratory}. We used Bayesian optimization with a random forest surrogate model to effectively search the parameter spaces with up to 6 million different configurations.  By using ytopt to identify the best configuration, we achieved up to 91.59\% performance improvement, up to 21.2\% energy savings, and up to 37.84\% EDP improvement on up to 4,096 nodes. 

Nevertheless, our current ytopt has several limitations. ytopt evaluates one parameter configuration each time. This affects the effectiveness of identifying the promising search regions at the beginning of the autotuning process because of few data points being available for training the random forest surrogate model. This one-by-one evaluation process is time-consuming. There is a need for ytopt to parallelize the multiple independent evaluations in order to accelerate the autotuning process. Some of the initial random selected configurations may result in much larger execution time than the baseline one as we had in \cite{R1}; there is also a need for ytopt to set a proper evaluation timeout in order to evaluate more good configurations. These are the main motivations for the work in this paper.

libEnsemble \cite{R14, R13}, which is part of the ECP PETSc/TAO project \cite{R45}, is a Python toolkit for coordinating workflows of asynchronous and dynamic ensembles of calculations across massively parallel resources.
libEnsemble helps users take advantage of massively parallel resources to solve design, decision, and inference problems and expands the class of problems that can benefit from increased parallelism. libEnsemble employs a manager/worker scheme that communicates via MPI, multiprocessing, or TCP. 

To cope with the limitations of ytopt mentioned earlier, based on our previous work on autotuning the performance, power, and energy of applications and systems \cite{R1, R2, R9}, we propose an asynchronously autotuning framework {\tt ytopt-libe} by integrating ytopt and libEnsemble. The framework {\tt ytopt-libe} comprises two asynchronous aspects:
\begin{itemize}
 \item The asynchronous aspect of the search allows the search to avoid waiting for all the evaluation results before proceeding to the next iteration. As soon as an evaluation is finished, the data is used to retrain the surrogate model, which is then used to bias the search toward the promising configurations.
\item The asynchronous aspect of the evaluation allows {\tt ytopt} to evaluate multiple selected parameter configurations in parallel by using the asynchronous and dynamic manager/worker scheme in {\tt libEnsemble}. All workers are independent; there is no direct communication among them.
When a worker finishes an evaluation, it sends its result back to  {\tt ytopt}. Then {\tt ytopt} updates the surrogate model and selects a new configuration for the worker to evaluate.
\end{itemize}

The advantages of this proposed autotuning framework {\tt ytopt-libe} are threefold:
\begin{itemize}
 \item Exploit the asynchronous and dynamic task management features provided by {\tt libEnsemble}
\item Accelerate the evaluation process of {\tt ytopt} to take advantage of massively parallel resources by overlapping multiple evaluations in parallel
\item Improve the accuracy of the random forest surrogate model by feeding more data to make the {\tt ytopt} search more efficient.
\end{itemize}

OpenMC~\cite{R11,R10} is a community developed Monte Carlo neutral transport code. Recent years have seen  development and optimization of this code for GPUs via the OpenMP target offloading model, as part of the ECP ExaSMR project \cite{R47}. In this paper we apply the proposed framework {\tt ytopt-libe} to autotune OpenMC to evaluate its effectiveness on an ECP system Crusher  \cite{R5}, which is a Frontier test and development system at the Oak Ridge Leadership Computing Facility (OLCF).

\xingfu{
This paper makes the following contributions.
 \begin{itemize}
 \item We extend and enhance the ytopt autotuning framework by leveraging the libEnsemble to propose an asynchronously autotuning framework ytopt-libe to tackle its limitations.
  \item We use this ytopt-libe framework to explore the tradeoffs between application runtime and power/energy for energy efficient application execution.
\item We apply this framework to autotune the very complex ECP application OpenMC, using Bayesian optimization with a Random Forest surrogate model to effectively search a parameter space with millions of configurations.
\item We demonstrate the effectiveness of the ytopt-libe framework to tune the performance, energy, and EDP of the OpenMC.
\end{itemize}
}

The remainder of this paper is organized as follows. Section 2 discusses the background, challenges, and motivation of this study. Section 3 proposes our autotuning framework for integrating ytopt and libEnsemble. Section 4 describes the OLCF Frontier TDS system Crusher and its power measurement.  Section 5 describes the ECP application OpenMC and defines its parameter space.  Section 6 illustrates autotuning in performance based on the figure of merit (FoM) of OpenMC. Section 7 presents autotuning in the runtime, energy, and energy delay product (EDP). Section 8 summarizes this paper. 

\section{Background}
In this section we briefly discuss some background about ytopt and libEnsemble.
\subsection{ytopt: ML-Based Autotuning Software}

ytopt \cite{R1, R2, R3}, which is one of ECP PROTEAS-TUNE project \cite{R44}, is a ML-based autotuning software package that consists of sampling a small number of input parameter configurations, evaluating them, and progressively fitting a surrogate model over the input-output space until exhausting the user-defined maximum number of evaluations or the wall-clock time. The package is built based on Bayesian optimization that solves optimization problems.

\begin{figure}[ht]
\center
 \includegraphics[width=\linewidth]{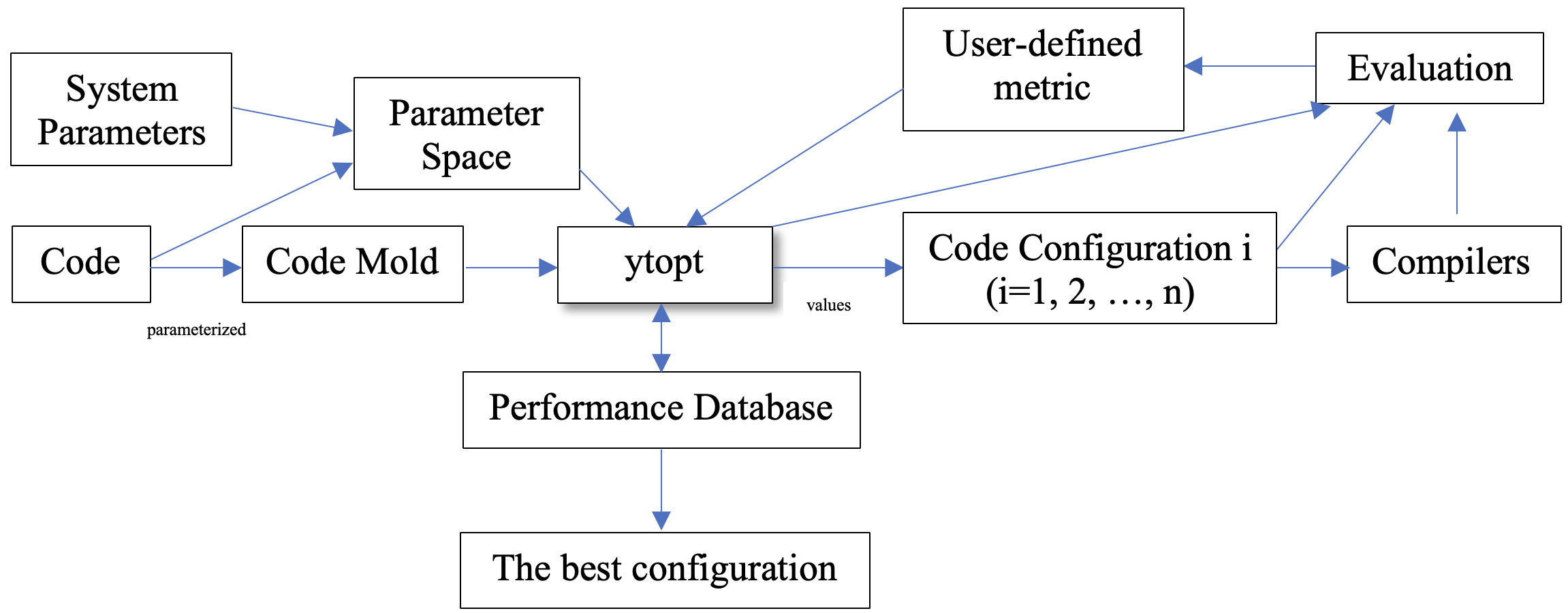}
 \caption{ytopt autotuning framework}
\label{fig:pf}
\end{figure}

Figure \ref{fig:pf} presents the framework for autotuning various applications. The application runtime is the primary user-defined metric. We use ConfigSpace to analyze an application code in order to identify the important tunable application and system parameters needed to define the parameter space using ConfigSpace \cite{R7}.
We use the tunable parameters to parameterize an application code as a code mold. 
ytopt starts with the user-defined parameter space, the code mold, and user-defined interface that specifies how to evaluate the code mold with a particular parameter configuration. 

The search method within ytopt uses Bayesian optimization, where a dynamically updated random forest surrogate model that learns the relationship between the configurations and the performance metric is used to balance exploration and exploitation of the search space. In the exploration phase, the search evaluates parameter configurations that improve the quality of the surrogate model. In the exploitation phase, the search evaluates parameter configurations that are closer to the previously found high-performing parameter configurations. The balance is achieved through the use of the lower confidence bound (LCB) acquisition function that uses the surrogate model's predicted values of the unevaluated parameter configurations and the corresponding uncertainty values. The LCB acquisition function is defined in Equation \ref{eqn:lcb}. For the unevaluated parameter configuration $x_M$, the trained model $M$ is used to predict a point estimate (mean value) $\mu(x_M)$ and standard deviation $\sigma(x_M)$.
\begin{equation}
    a_{LCB}(x_M) = \mu(x_M) - \kappa\sigma(x_M),
    \label{eqn:lcb}
\end{equation}
where $\kappa \geq 0$ is a user-defined parameter that controls the tradeoff between exploration and exploitation. When $\kappa=0$ for pure exploitation, a configuration with the lowest mean value is selected. When $\kappa$ is set to a large value ($>1.96$) for pure exploration, a configuration with large predictive variance is selected. The default value of $\kappa$  is 1.96. Then the model $M$ is updated with this selected configuration.

In our previous work \cite{R1} we autotuned four hybrid MPI/OpenMP ECP proxy applications---XSBench, AMG, SWFFT, and SW4lite---at large scales. We used Bayesian optimization with a random forest surrogate model to effectively search the parameter spaces with up to 6 million different configurations on Theta \cite{R19} at Argonne National Laboratory and Summit \cite{R20} at the Oak Ridge Leadership Compouting Facility (OLCF). The experimental results showed that our autotuning framework had low overhead and good scalability. By using ytopt to identify the best configuration, we achieved up to 91.59\% performance improvement, up to 21.2\% energy savings, and up to 37.84\% EDP improvement on up to 4,096 nodes. 

Nevertheless, our current ytopt has several limitations. ytopt evaluates one parameter configuration each time. This reduces the effectiveness of identifying the promising search regions at the beginning of the autotuning process because of having few data points for training the random forest surrogate model. This one-by-one evaluation process is also time-consuming. There is a need for ytopt to parallelize the multiple independent evaluations to accelerate the autotuning process. Some of initial randomly selected configurations may result in much larger execution time than the baseline as we had in \cite{R1}. There is also a need for ytopt to set a proper evaluation timeout in order to evaluate more good configurations. These are the main motivations for this work.

\subsection{libEnsemble} 

libEnsemble \cite{R14, R13} is an advanced Python toolkit, developed under the ECP PETSc/TAO project \cite{R45}, designed to coordinate workflows of asynchronous, dynamic ensembles of calculations on massively parallel resources. Targeting complex design, decision-making, and inference problems, libEnsemble allows applications and studies where single instances typically do not scale to an entire machine to effectively harness high-performance computing environments.

\begin{figure}[ht]
    \centering
    \includegraphics[width=\linewidth]{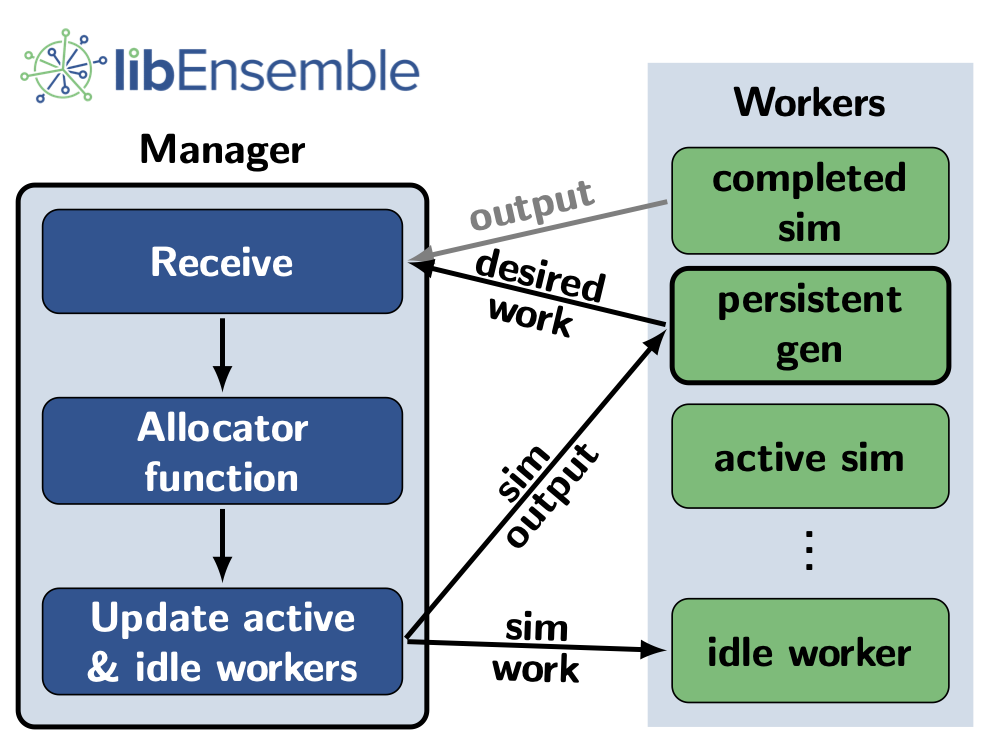}
    \caption{libEnsemble's manager-worker paradigm: the manager allocates workload to multiple workers, which perform computations via user functions}
    \label{fig:libe}
\end{figure}

One of the key strengths of libEnsemble is its extreme portability and scalability. It operates seamlessly across various computing platforms, from laptops to clusters to leadership-class machines. This flexibility ensures that the toolkit can be employed in a wide range of environments without extensive configuration or modification. Additionally, libEnsemble supports heterogeneous computing workflows, dynamically and portably assigning and reassigning CPUs, GPUs, or multiple nodes to tasks, making it an ideal solution for workflows using diverse computational resources.

libEnsemble employs a manager-worker paradigm, depicted in Figure \ref{fig:libe}, that significantly enhances its capability to handle dynamic and asynchronous calculation workflows. The manager and workers together coordinate three core plug-and-play components: simulation (sim) functions, generator (gen) functions, and allocator functions, all referred to as \textit{user functions}. The simulation function is responsible for executing the individual calculations within the ensemble. These could range from complex simulations to simpler data-processing tasks. The generator function plays a crucial role in deciding the next set of calculations to be performed, often based on the results of previous simulations. Generators can adjust the direction of the computational workflow dynamically, ensuring that the ensemble is always computing on the most relevant and promising tasks. The allocator function decides what tasks are given to which workers based on the current state of the ensemble and the available resources. This modular approach allows libEnsemble to dynamically adjust its focus based on intermediate results. Such flexibility is key to efficiently navigate large, complex computational spaces.

libEnsemble also includes advanced features to automatically monitor running calculations and manage data-flow between tasks, which enhances libEnsemble's utility in complex studies where intertask communication is crucial. At a baseline, simple send and receive methods are available to ensemble members to send intermediate data and hyperparameters to each-other via the manager. From this foundation, a critical capability includes cancelling or preempting application instances, as directed by an autotuning or other decision process, providing scientists with more efficient control over the study.

Another significant advantage of libEnsemble is its low start-up cost for domain scientists, regardless of their familiarity with parallel computing. It requires no additional background services or processes to function, making it straightforward to integrate and deploy in various computational environments. This ease of setup, combined with its powerful features, positions libEnsemble as a highly versatile and user-friendly toolkit for coordinating workflows in HPC environments. The complete documentation of libEnsemble is presented in \cite{R15}; examples of diverse applications of libEnsemble are demonstrated in \cite{R16,R17,R18}.

\section{Integrated Autotuning Framework ytopt-libe in Performance and Energy}

In this section we discuss the integration of the ytopt autotuning framework and libEnsemble to propose a new autotuning framework ytopt-libe using a user-defined metric such as the application FoM, application runtime, energy, or EDP, where the application FoM 
is the primary performance metric. Energy consumption captures the tradeoff between the application runtime and power consumption; and EDP captures the tradeoff between the application runtime and energy consumption. 

\begin{figure}[ht]
\center
  \includegraphics[width=\linewidth]{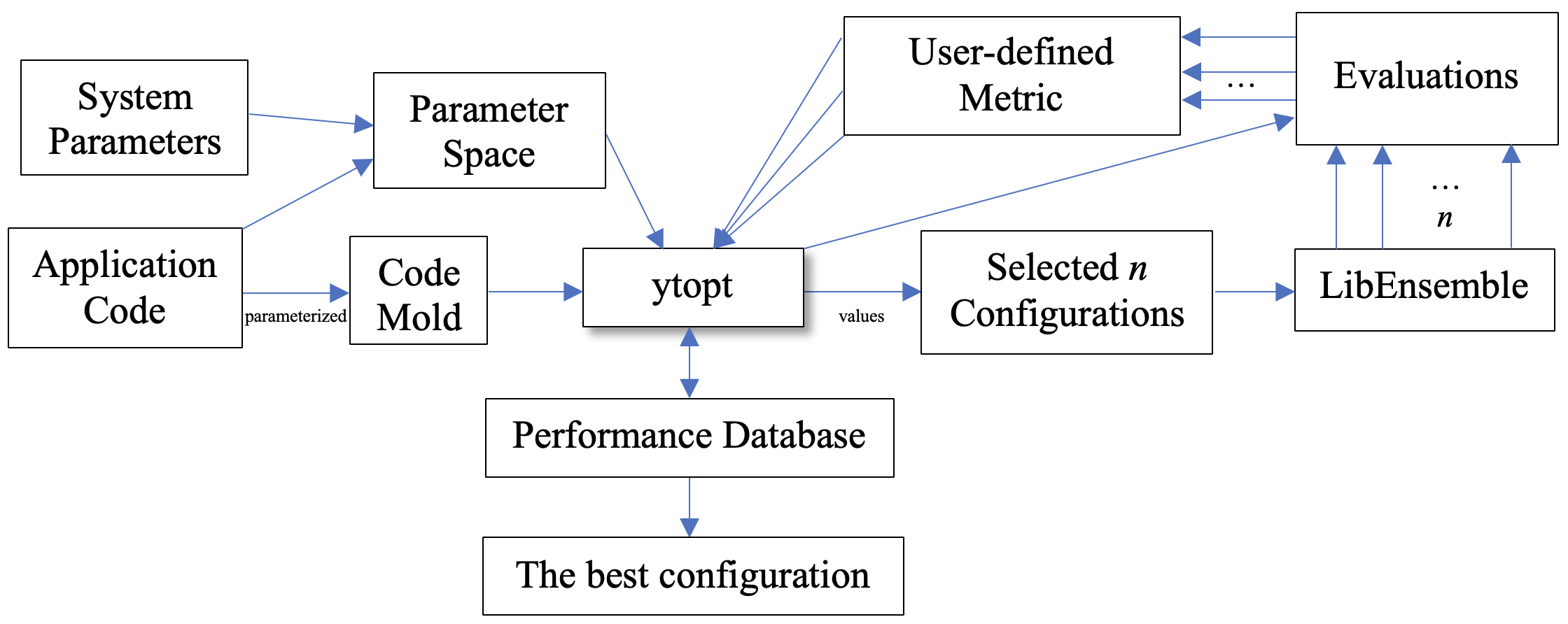}
 \caption{Integrated autotuning framework ytopt-libe}
\label{fig:yl}
\end{figure}

Figure \ref{fig:yl} presents the framework ytopt-libe for autotuning various applications in user-defined metrics. 
We analyze an application code to identify the important tunable application parameters and system parameters (OpenMP environment variables and number of MPI ranks per GPU) to define the parameter space using ConfigSpace.
We use these tunable parameters to parameterize an application code/script as a code mold. 
ytopt-libe starts with the user-defined parameter space, the code mold, and user-defined interface that specifies how to evaluate the code mold with a particular parameter configuration. 

For the integration of ytopt and LibEnsemble, the generator function in libEnsemble uses the ytopt's Bayesian optimization with random forest surrogate models to drive the ensemble of calculations, the simulation function is the evaluation of the ExaSMR application calculation, and the allocator function uses the libEnsemble default for a persistent worker operating in batch mode. Upon initialization, a worker is dedicated to running the persistent generator. For this case, batch behavior is desired, so the allocator function tells the generator to produce a point to evaluate for each of the remaining idle workers. The allocator function dispatches these points out; and once all of these simulation runs are completed, the persistent generator is informed of the results, and the process is repeated.

The iterative phase of the proposed autotuning framework ytopt-libe has the following steps: 
\begin{itemize}
\item [Step 0]  libEnsemble creates one manager and $n$ independent workers specified by a user.
\item [Step 1]  Bayesian optimization selects $n$ parameter configurations for evaluation.
\item [Step 2] The code mold is configured with each selected configuration to generate new code for each worker in parallel.
\item [Step 3]  For each worker, based on the number of threads in each configuration, the number of nodes reserved, and the number of MPI ranks, the job command line using srun/mpirun for the launch of the application on the compute nodes is generated in parallel.
\item [Step 4] Each new code is compiled with other codes needed to generate an executable in parallel for each worker (if needed).
\item [Step 5] For each worker, the generated command line is executed to evaluate the application with the selected parameter configuration in parallel; each result is sent back to the ytopt and is recorded in the performance database in an asynchronous and dynamic way: 
(1)	All workers are independent; there is no communication among them.
(2)	When a worker finishes an evaluation, it sends its result back to ytopt; then ytopt updates the surrogate model and selects a new configuration for the worker to evaluate (Step 1).
\end{itemize}

Steps 1--5 are repeated until the maximum number of code evaluations or the wall-clock time is exhausted for the autotuning run. 

In this way, the proposed autotuning framework ytopt-libe not only accelerates the evaluation process of ytopt to take advantage of massively parallel resources but also improves the accuracy of the random forest surrogate model by feeding more data for more efficient search. 

 To set a proper evaluation timeout in order to evaluate more good configurations,  ytopt requires the timeout value as a user input. The basic principle is based on the baseline application runtime plus some overhead. Therefore, the typical timeout setting is 1.5 times the baseline runtime. ytopt uses Python \textit{subprocess.Popen.communicate} \cite{R22} to terminate an evaluation process after the timeout and sets the timeout as the returned objective value for ytopt to record the value and the configuration in the performance database. Then ytopt searches the next configuration for evaluation.

\section{OLCF System Crusher} 

The Crusher \cite{R5} system has identical hardware and similar software to those of the Frontier system at OLCF. Crusher has 2 cabinets, the first with 128 compute nodes and the second with 64 compute nodes, for a total of 192 compute nodes. 
\begin{figure}[ht!]
\center
 \includegraphics[width=\linewidth]{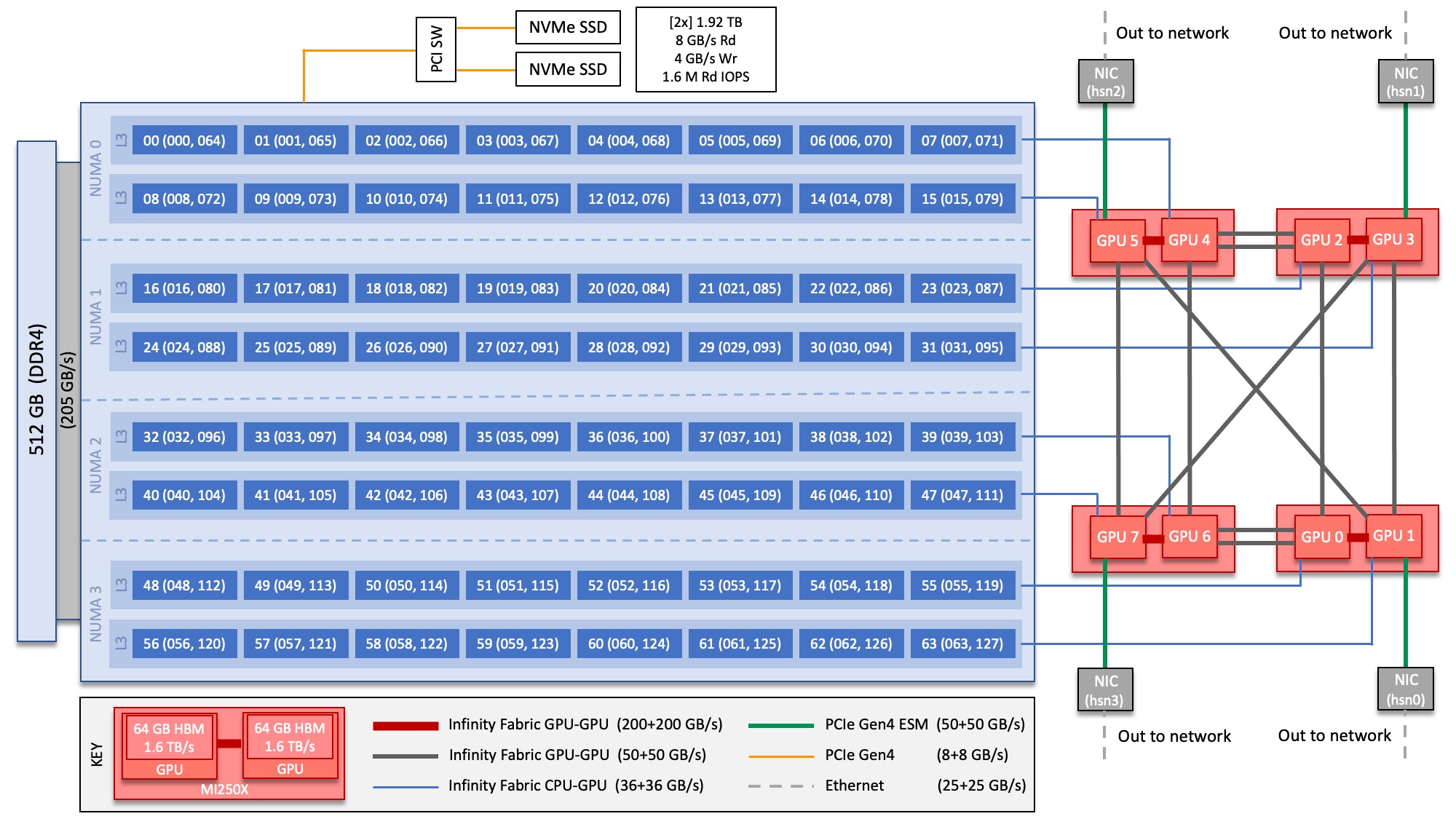}
 \caption{Crusher node diagram \cite{R5}.}
\label{fig:sys}
\end{figure}

As shown in Figure \ref{fig:sys},  each Crusher compute node consists of one 64-core AMD EPYC 7A53 “Optimized 3rd Gen EPYC” CPU with 2 hardware threads per core with access to 512 GB of DDR4 memory. Each node also contains four AMD MI250X, each with 2 Graphics Compute Dies (GCDs) for a total of 8 GCDs per node. The CPU is connected to each GCD via Infinity Fabric CPU-GPU, allowing a peak host-to-device  and device-to-host bandwidth of 36+36 GB/s. The 2 GCDs on the same MI250X are connected with Infinity Fabric GPU-GPU with a peak bandwidth of 200 GB/s. The GCDs on different MI250X are connected with Infinity Fabric GPU-GPU, where the peak bandwidth ranges from 50 to 100 GB/s based on the number of Infinity Fabric connections between individual GCDs. 
We  refer to the GCDs simply as GPUs in this paper. Slurm \cite{R48} is the workload manager used to interact with the compute nodes on Crusher; srun is used to launch application jobs on Crusher’s compute nodes in Slurm.

\begin{figure}[ht]
\center
 \includegraphics[width=\linewidth]{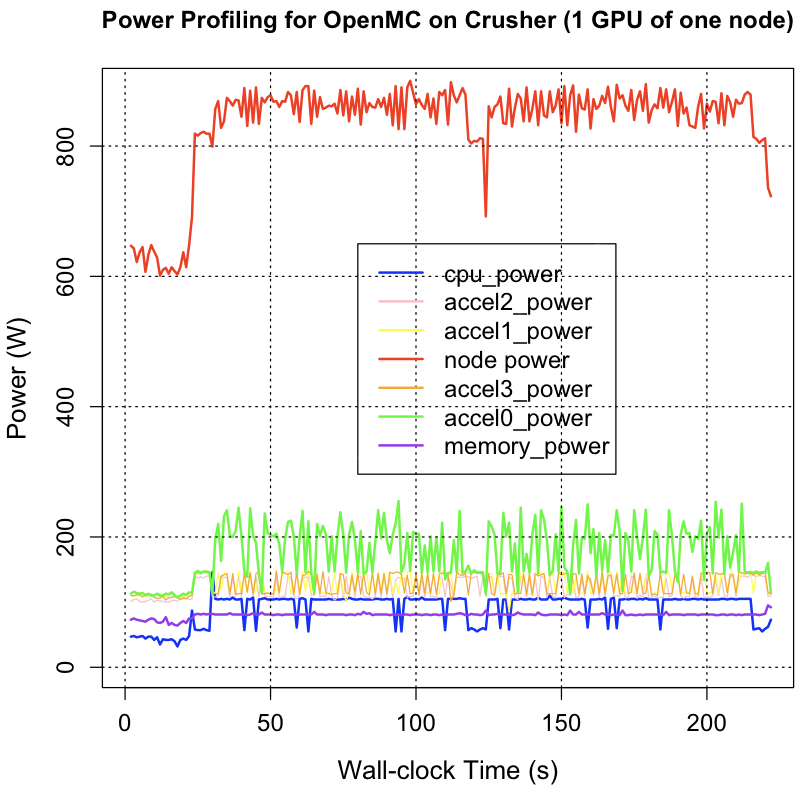}
 \caption{Power profiling for running OpenMC on a single GPU on Crusher}
\label{fig:p1}
\end{figure}

We use APEX \cite{R8} to read the Cray pm\_counters to obtain the power consumptions for each node and its components, such as CPU, GPUs, and memory. For instance, Figure \ref{fig:p1} shows the power profiling over time for OpenMC with 2 iterations running on one GPU of a node using APEX, where one of the GPUs from AMD MI250X accelerator 0 was used for running the OpenMC and other accelerators are idle; there are also some activities in the CPU and a few activities in memory for this application. Node power is the overall power consumption for the node.

\section{ECP Application OpenMC and Its Parameter Space}
\subsection{ECP Application OpenMC} %

OpenMC~\cite{R11} is a community-developed Monte Carlo neutral transport code. Recent years have seen  development and optimization of the code for GPUs via the OpenMP target offloading model, as part of the ECP ExaSMR project \cite{R47}. The goal of the ExaSMR project was to simulate small modular nuclear reactors in high fidelity using a multiphysics approach that coupled Monte Carlo neutron transport with computational fluid dynamics. OpenMC has so far demonstrated excellent performance across a variety of GPU architectures, including Intel, NVIDIA, and AMD~\cite{R10}. While OpenMC's portability is a great benefit for its users, however, fully fine-tuning the code for each GPU architecture by hand is impractical. Rather, an automated approach is desired so as to find optimal parameters for each new system and to determine whether they tend to differ between systems or whether a common default can be used everywhere.

In this study we  perform tuning and analysis with OpenMC on a standard reactor physics benchmark simulation problem---the Hoogenboom--Martin (HM) ``large'' variant~\cite{R12} representing a depleted fuel reactor with 272 nuclides in total. Optimal performance parameters discovered when tuning for this problem are expected to translate directly to a wide variety of fission reactor simulation problems, including the ExaSMR challenge problem. 

\begin{table*}[ht]
\small\sf\centering
\caption{OpenMC tunable parameters}
\begin{tabular}{llll}
\toprule
Parameter & Type & Range/Bounds & Default \\
\midrule
Maximum number of particles in-flight &  Integer & 100k - max that will fit in memory ($\sim$ 8 million)  & 1 million \\
Number of MPI ranks per GPU &Integer  & 1 - 4 &  1 \\
Number of logarithmic hash grid bins  &Integer & 100 - 100k &  4,000 \\
Queuing logic type &Mode  & Queued vs. Queueless &  Queued \\
Minimum sorting threshold (queued mode only) &Integer &0 (always sort) - infinity (never sort)  & 20,000 \\
\bottomrule
\end{tabular}
\label{tab:tp}
\end{table*}

The tunable parameters and their ranges of OpenMC with the HM Large benchmark problem are shown in Table \ref{tab:tp}. A description of each parameter is given below.

\begin{itemize}

\item The {\bf maximum number of particles in-flight} parameter controls the number of particles that are allowed to be alive at any one point in time. This parameter does not affect the numerical characteristics of the simulation, since the user is still allowed to select any number of particle histories per simulation batch. If the number of particles allowed in-flight is below the number of particles per batch selected by the user, then the particle buffer will be refilled dynamically after each kernel call to replace any particles that have reached their termination criteria, until all particle histories have been simulated in the batch. The motivating reason for this parameter is  both to control memory usage (since particle objects are relatively large) and to allow for more optimal kernel work sizes to be selected. Based on our experience, the maximum number of particles in-flight is in the range of 100,000 and $\sim$ 8 million, with the default of 1 million.

\item The number of {\bf MPI ranks per GPU} parameter is useful since some architectures see significant performance gains when multiple MPI ranks are launched per GPU, in effect resulting in multiple streams running on the GPU concurrently. This variable is correlated with others in that selection of a higher number of MPI ranks will greatly reduce the memory available to each rank, which constrains selection of other parameters that involve an increased usage of memory. We need to tune the parameter to identify the proper number of MPI ranks per GPU for the best performance.

\item The number of {\bf logarithmic hash grid bins} is a tunable parameter that affects performance of the most expensive kernel in OpenMC for depleted nuclear reactor simulations---the macroscopic cross-section lookup kernel. This parameter determines how many logarithmic hash bins the energy space is divided into. If only a single bin were to be selected, then each nuclide in a given material in the simulation would need to be searched for data that corresponds to the neutron's current energy at each stage of the simulation. For depleted fuel materials, this results in hundreds of separate binary search operations per macroscopic lookup. If a very high number of hash bins are selected, then all search operations can be eliminated and replaced with direct access operations. While a choice of a very high number of grid bins is often much faster, it comes at the cost of impractically high memory usage requirements. Thus, there is an inherent balance between memory usage and speedup that can result in architecture-specific tuning of this parameter. Choices resulting in high memory usage may also reduce the number of particles in-flight that are possible, making this variable somewhat correlated to others that result in more memory pressure.

\item The {\bf queuing logic} parameter selects between two different control flow implementations. The queued method involves the host maintaining discrete queues of particles representing each event type in the simulation. The host selects which kernel to launch on the device based on which queue has the most particles in it. This sort of dynamic queuing is useful given the stochastic nature of the Monte Carlo transport process. A result of this policy is that the number of particles in each queue must be transmitted back to the host, potentially increasing kernel invocation overhead and creating bubbles when kernel runtimes are very low. The second option, the queueless mode, uses simpler logic on the host that just loops over all event types until all particles have terminated. This eliminates the need to transfer any particle count data between the host and the device after each kernel invocation, potentially reducing overhead, but at the loss of having to subscribe threads for particles that do not need that event. Such threads simply check whether their particle requires the event and return immediately if the kernel is not needed, potentially resulting in inefficient thread divergence.

\item The {\bf minimum sorting threshold} parameter sets a limit on the number of particles that must be present in a queue to undergo sorting. Below this limit, no sorting is performed. The sorting operation improves memory locality of cross-section lookup operations by ensuring that adjacent particles in the buffer are located in the same material and close in energy. If many particles are in-flight and sorting is performed, potentially all threads within a workgroup (block) may be accessing the same data. However, if there are very few particles in-flight, then sorting may not result in any meaningful improvements in locality because threads may be accessing data that is no longer on the same cache line. In this case, the cost of the sorting operation may outweigh the benefits, so a minimum threshold parameter is useful for tuning.

\end{itemize}

\subsection{Parameter Space for OpenMC}

Based on the tunable application parameters in Table \ref{tab:tp} and the system parameters such as the number of threads and OpenMP thread placement, we 
define the parameter space for OpenMC using ConfigSpace as follows.

{\scriptsize
\begin{verbatim}
cs = CS.ConfigurationSpace(seed=1234)
# queuing logic type 
P0 = CSH.CategoricalHyperparameter(name='P0', choices=
["openmc", "openmc-queueless"], default_value="openmc")
# maximum number of particles in-flight (option -i)
P1= CSH.UniformIntegerHyperparameter(name='P1', lower=100000, 
upper=8000000, default_value=1000000, q=1000)
# number of logarithmic hash grid bins (option -b)
P2 = CSH.UniformIntegerHyperparameter(name='P2', lower=100, 
upper=100000, default_value=4000, q=100)
# minimum sorting threshold (option -m)
P3 = CSH.UniformIntegerHyperparameter(name='P3', lower=0, 
upper=1000000, default_value=20000, q=1000)
#number of threads
P4= CSH.UniformIntegerHyperparameter(name='P4', lower=2, 
upper=8, default_value=8)
#number of tasks per gpu
P5= CSH.OrdinalHyperparameter(name='P5',sequence=[1, 2], 
default_value=1)
# omp placement
P6= CSH.CategoricalHyperparameter(name='P6', choices=
['cores','threads','sockets'], default_value='threads')
cs.add_hyperparameters([P0, P1, P2, P3, P4, P5, P6])

cond = EqualsCondition(P3, P0, "openmc")
cs.add_conditions([cond])
\end{verbatim}
}  
Here P0 is the queuing logic type parameter with the choices of ``openmc'' (queued OpenMC code) and ``openmc-queueless'' (queueless OpenMC code); P1 is the maximum number of particles in-flight with the command line option -i; P2 is the number of logarithmic hash grid bins with the command line option -b; P3 is the minimum sorting threshold for only queued OpenMC code with the command line option -m; P4 is the number of OpenMP threads per task; P5 is the number of MPI tasks per GPU; and P6 is the OpenMP thread placement type, which may impact the OpenMP performance. The default values are set based on the application and system default settings, so the default configuration is the starting point for autotuning. 

Notice that there is some dependency among these parameters. Because P3 is only valid when P0 is the queued OpenMC, this constraint,
the ConfigSpace condition statement {\it{EqualsCondition(P3, P0,  ``openmc'')}} is added to the parameter space, which means that making P3 is an active hyperparameter if P0 has the value ``openmc''; otherwise, P3 has the empty value "nan".  P4 depends on P5 because each core of a 64-core AMD EPYC CPU supports two hardware threads, with a total of 128 threads per node. Using 2 MPI tasks per GPU means 16 MPI tasks per node, so  the number of threads P4 is at most 8.

In this paper, we use the hybrid MPI/OpenMP offload OpenMC code written in C++, MPI, and OpenMP offloading (version: June 10, 2022) which is very complex; its compiling time is very long. Therefore, to avoid the time-consuming compiling, we precompile OpenMC to two executables: openmc for the queued OpenMC and openmc-queueless for the queueless OpenMC. Because OpenMC provides the command line options for these application parameters, we can avoid the code compiling and just pass these parameters to the OpenMC executable command line. 
Hence we parameterize the OpenMC command line and its options and define the code mold (openmc.sh) as follows.

{\scriptsize
\begin{verbatim}
#!/bin/bash 
pp0="#P0"
pp="openmc"
if [ "$pp0" = "$pp" ]
then
        openmc  --event -i #P1 -b #P2 -m #P3
else
        openmc-queueless --event -i #P1 -b #P2
fi
\end{verbatim}
}  
Here OpenMC uses event-based parallelism with the option "--event" for both queued and queueless modes. P3 is only for the queued OpenMC.
 
We also pass the parameters P4, P5, and P6 to the srun command with the following command line options.
{\scriptsize
\begin{verbatim}
-c #P4
--ntasks-per-gpu=#P5
--cpu-bind=#P6
\end{verbatim}
}  

After ytopt-libe selects a configuration, the parameters from P0 to P6 in the code mold and the srun command line are automatically replaced with the selected values in the configuration to generate a new run script for evaluation.

\section{Autotuning in Performance}

In this section we apply the proposed framework ytopt-libe, shown in  in Figure~\ref{fig:yl}, to autotune the performance of OpenMC. The default performance metric for OpenMC is the figure of merit, which is the rate to process the number of particles per second. Slurm's srun is used to launch an application to compute nodes on OLCF Crusher. The AMD EPYC CPU processor core for each Crusher node supports the simultaneous multithreading level of 2 as default so that the number of threads per node is supported up to 128. 

As shown in Figure~\ref{fig:yl}, ytopt-libe creates one manager and $n$ independent workers specified by a user, and Bayesian optimization in ytopt-libe selects $n$ parameter configurations for evaluation. Based on the number of threads from the selected configuration, the number of nodes reserved, and the number of MPI ranks, ytopt-libe generates the srun command line for application launch on compute nodes. 

For the experimental setup, we set the total number of evaluations 256 for all experiments and run each worker on the dedicated node(s). We use the following terms in the rest of this paper. $(1 GPU, 1 node)$ stands for using 1 GPU on a single node.  $(8 GPUs, 1 node)$ stands for using 8 GPUs on a single node. $(2 workers, 1 GPU/n) $ means that each worker runs on 1 GPU on a single dedicated node with the total 2 nodes used for 2 workers.  $(4 workers, 8 GPUs/n)$ means that each worker runs on 8 GPUs on a single dedicated node with the total 4 nodes used for 4 workers. $(4 workers, 8 nodes)$ means that each worker runs on 2 nodes with 8 GPUs/node and the total 8 nodes for 4 workers. Notice that each worker is assigned to one or more dedicated nodes so that its performance is not impacted by other workers.

\xingfu{The baseline performance measurement is based on the default values for application parameters for OpenMC with both queueless and queued modes on Crusher with the default system environment settings. We use OpenMC with the HM-Large benchmark problem for all our experiments. We run both modes (queueless, queued) with the default application settings from Table \ref{tab:tp} on the default system environment (1 MPI rank per GPU, 8 threads per CPU, omp placement: threads) of Crusher five times and use the largest particles/s as the baseline for performance comparison.}

\subsection{Performance Comparison}
In this subsection we compare the performance using ytopt itself and the proposed framework ytopt-libe in the following two ways:
\begin{itemize}
 \item Investigate how ytopt-libe accelerate the autotuning evaluations for ytopt. 
\item Explore a proper number of workers to be used for identifying the best configuration (varying the number of workers from 2 to 16) using ytopt-libe. 
\end{itemize}

\begin{figure}[ht]
\center
  \includegraphics[width=\linewidth]{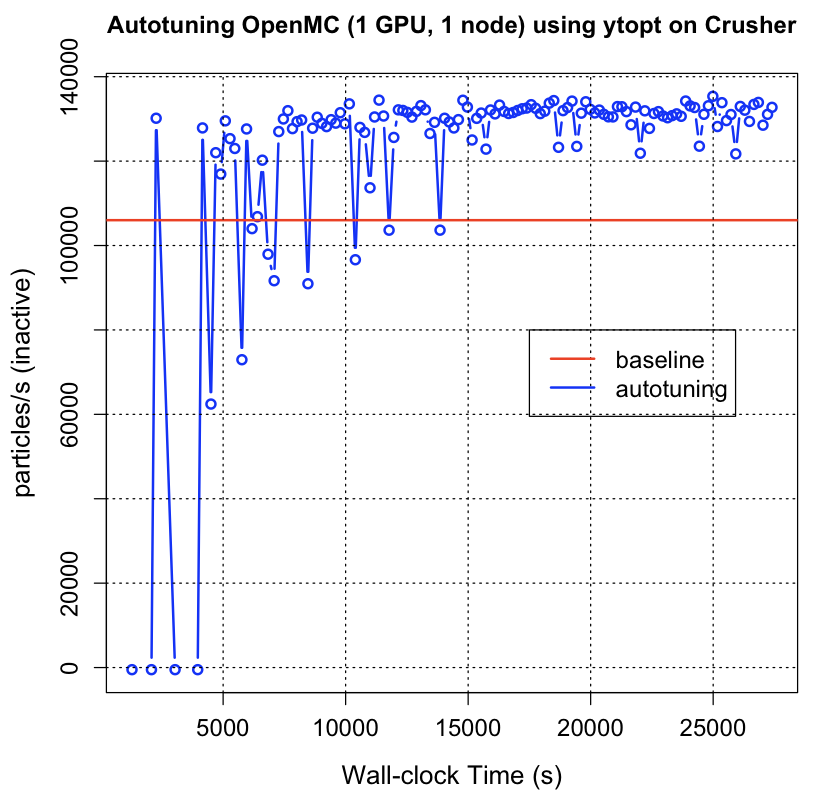}
 \caption{Autotuning OpenMC on a single GPU using ytopt}
\label{fig:yt1}
\end{figure}

Figure \ref{fig:yt1} shows autotuning OpenMC over time on a single GPU using ytopt, where the red line stands for the baseline performance and the blue circle is the FoM for a selected configuration evaluation. Over  time, ytopt leads to the high-performing regions of the parameter space, which result in much better performance than the baseline performance. However, several initial configurations had a timeout termination so that its performance was set very low with a constant negative value by ytopt. In this way, the timeout configurations were marked as the worst regions so that the next search would avoid these regions. Using ytopt took 27,401 s to finish just 128 of 256 evaluations because of the large application runtimes (more than 200 s). ytopt identified the best configuration, which resulted in 27.68\% performance improvement.

\begin{figure}[ht]
\center
  \includegraphics[width=\linewidth]{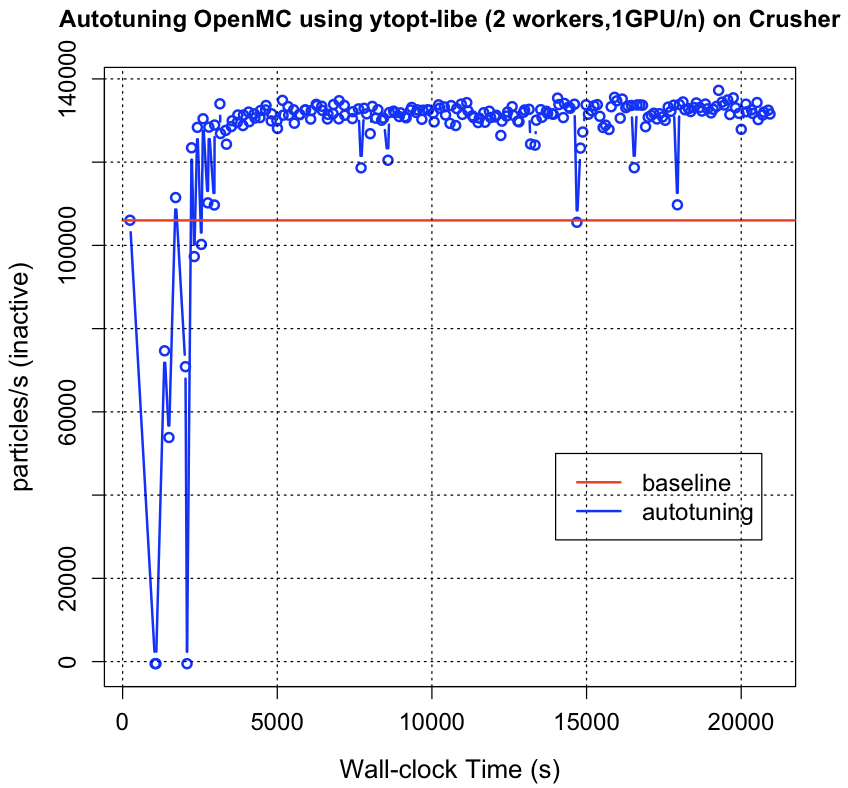}
 \caption{Autotuning OpenMC on a single GPU using ytopt-libe with 2 workers}
\label{fig:yl1-2}
\end{figure}

When ytopt-libe with 2 workers on two nodes is used to autotune OpenMC on a single GPU, as shown  in Figure \ref{fig:yl1-2}, ytopt-libe creates two independent workers and selects two configurations: one worker runs the OpenMC with one configuration on a single GPU of one node, and the other worker runs the OpenMC with the other configuration on a single GPU of another node. In this way, two evaluations are executed independently in parallel. It took 20,931 s to finish 256 evaluations, reducing the overall autotuning time for using ytopt by more than a half. ytopt-libe identified the best configuration, which resulted in 29.48\% performance improvement.  

Can we use as many workers as possible for ytopt-libe to autotune OpenMC? To answer this question, we vary the number of workers from 2 to 16 to 
explore what is the best number of workers to use for identifying the best configuration. 

\begin{figure}[ht]
\center
  \includegraphics[width=\linewidth]{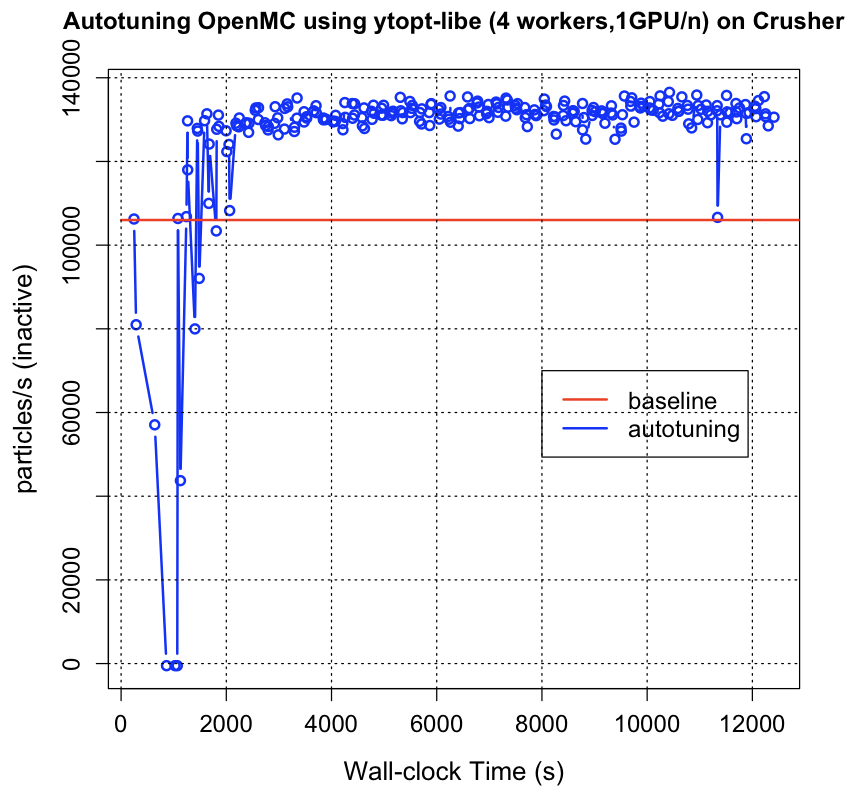}
 \caption{Autotuning OpenMC on a single GPU using ytopt-libe with 4 workers}
\label{fig:yl1-4}
\end{figure}

\begin{figure}[ht]
\center
  \includegraphics[width=\linewidth]{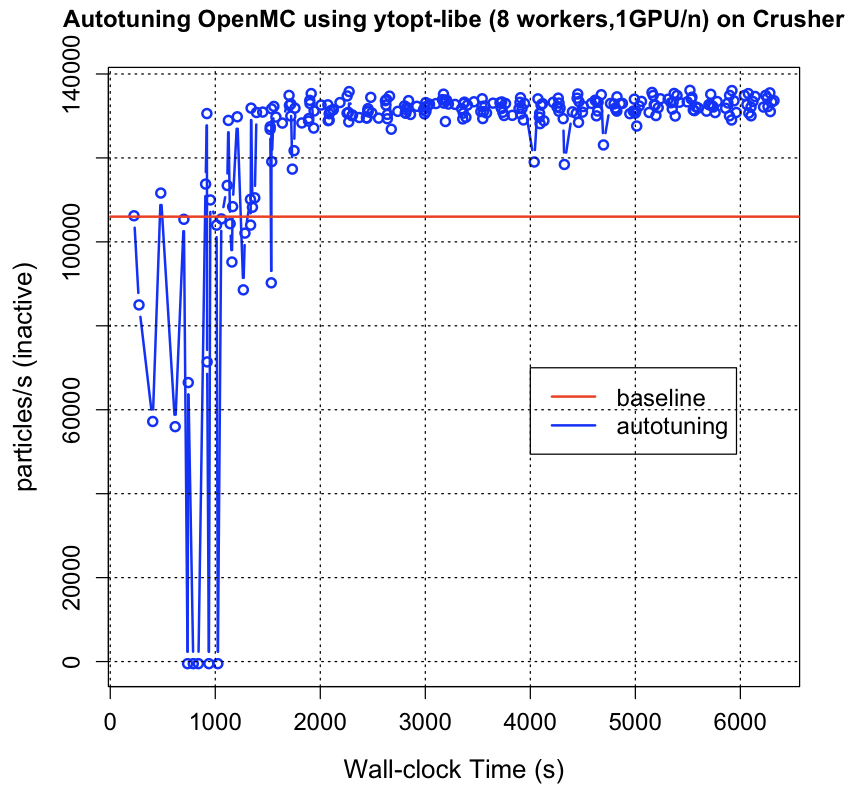}
 \caption{Autotuning OpenMC on a single GPU using ytopt-libe with 8 workers}
\label{fig:yl1-8}
\end{figure}

\begin{figure}[ht]
\center
  \includegraphics[width=\linewidth]{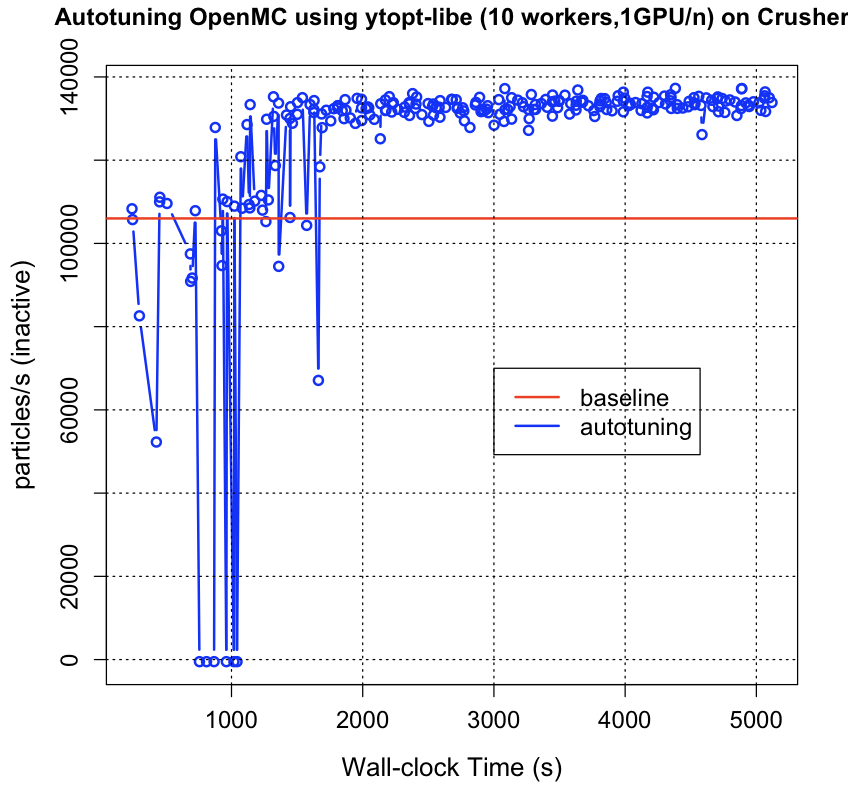}
 \caption{Autotuning OpenMC on a single GPU using ytopt-libe with 10 workers}
\label{fig:yl1-10}
\end{figure}

\begin{figure}[ht]
\center
  \includegraphics[width=\linewidth]{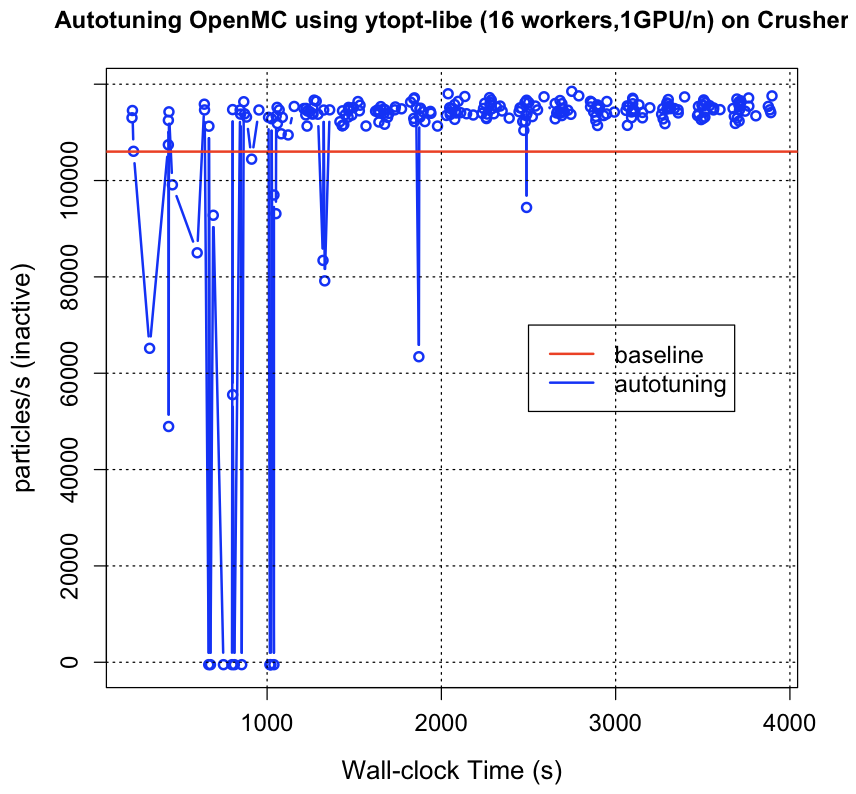}
 \caption{Autotuning OpenMC on a single GPU using ytopt-libe with 16 workers}
\label{fig:yl1-16}
\end{figure}

\begin{table*}[ht]
\small\sf\centering
\caption{Performance improvement comparison (based on the baseline) using ytopt and ytopt-libe with different number of workers}
\begin{tabular}{llll}
\toprule
Method & Number of evaluations & Performance improvement (\%) & Total autotuning time (s) \\
\midrule
ytopt &  128 of 256  & 27.68  & 27,401 \\
ytopt-libe with 2 workers & 256 & 29.48  &  20,931\\
ytopt-libe with 4 workers & 256 & 28.79  & 12,411\\
ytopt-libe with 8 workers & 256 & 28.39  & 6,320\\
ytopt-libe with 10 workers & 256 & 29.49  & 5,119\\
ytopt-libe with 16 workers & 256 & 11.81 & 3,897\\
\bottomrule
\end{tabular}
\label{tab:ytlc}
\end{table*}

Similarly, to autotune OpenMC on a single GPU per node with a total of 256 evaluations, we conduct the following experiments. We configure ytopt-libe  to use 4 workers on 4 nodes in Figure \ref{fig:yl1-4},  8 workers on 8 nodes in Figure \ref{fig:yl1-8}, 10 workers on 10 nodes in Figure \ref{fig:yl1-10}, and 16 workers on 16 nodes in Figure \ref{fig:yl1-16}. Table \ref{tab:ytlc} summarizes the performance improvement comparison using ytopt and ytopt-libe with different numbers of workers on Crusher. We observe that ytopt-libe significantly reduces the time spent in the ytopt autotuning process and that ytopt-libe with 10 workers resulted in the best performance improvement, 29.49\%. Using larger numbers of workers, however, may not necessarily achieve the best performance improvement. Especially for the case of ytopt-libe with 16 workers, the performance improvement significantly decreases to just 11.81\%. The reason is as follows. When ytopt-libe used 16 workers, it selected 16 configurations to start evaluation at the beginning of the autotuning process. Unfortunately, more bad configurations were selected for the evaluation, as shown in Figure \ref{fig:yl1-16}. These data points were fed to the random forest surrogate model, causing a less accurate model. This results in an inefficient search for the next configurations. Therefore, we have to deal with the trade-off among the runtime, performance improvement, and  number of available compute resources. 

\begin{figure}[ht]
\center
  \includegraphics[width=\linewidth]{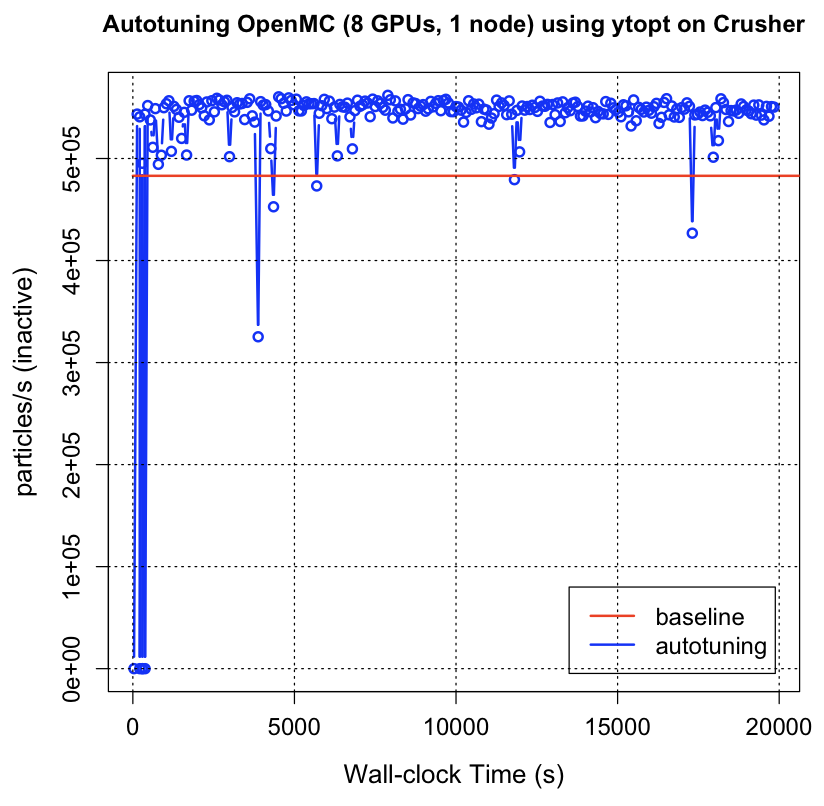}
 \caption{Autotuning OpenMC on 8 GPUs using ytopt}
\label{fig:yt8}
\end{figure}

\begin{figure}[ht]
\center
  \includegraphics[width=\linewidth]{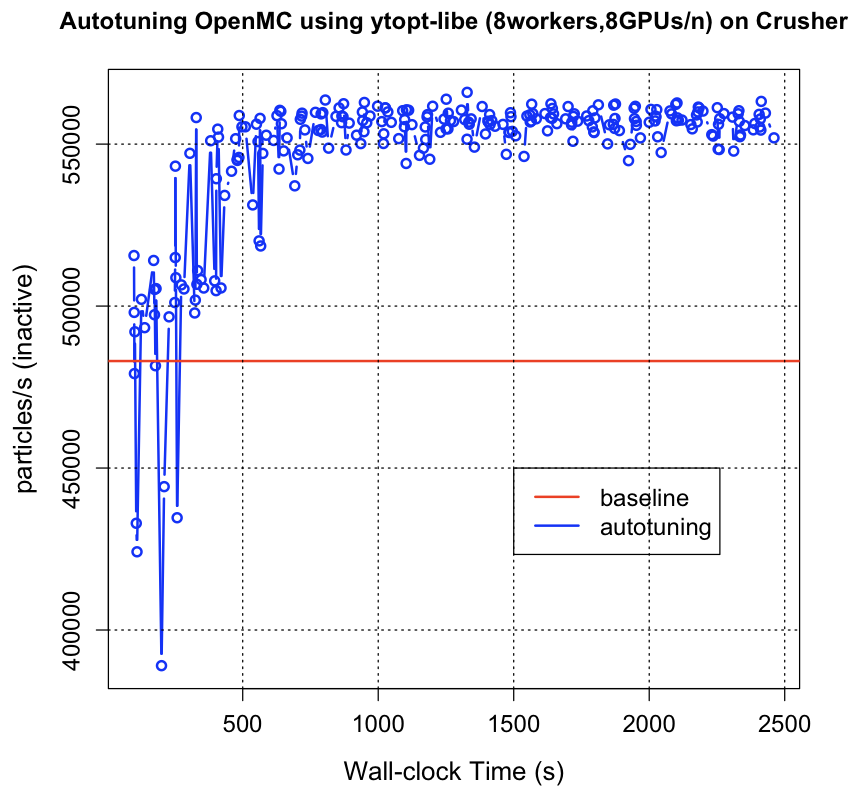}
 \caption{Autotuning OpenMC on 8 GPUs using ytopt-libe}
\label{fig:yl8-8}
\end{figure}

Figure \ref{fig:yt8} shows autotuning OpenMC over time on 8 GPUs on a single node using ytopt. Over time, ytopt leads to the high-performing regions of the parameter space, which results in much better performance than the baseline performance. It took 19,631 s to finish 256 evaluations. ytopt identified the best configuration, which resulted in 16.33\% performance improvement. Notice that using 8 GPUs per node for OpenMC resulted in much higher FoM than using 1 GPU per node in Figure \ref{fig:yt1}. Figure \ref{fig:yl8-8} shows autotuning OpenMC over time on 8 GPUs of a single node using ytopt-libe with 8 workers. This significantly reduced the total ytopt autotuning time by almost eight times and also achieved 17.17\% performance improvement. Using ytopt-libe with different numbers of workers for the 8 GPUs per node cases, we found that the trend was similar to that for the 1 GPU per node cases, as shown in Table \ref{tab:ytlc}.

\subsection{Performance Scaling Analysis}
In this subsection we investigate the application performance scaling using ytopt-libe. For the  analysis, we use a constant number of workers,  4,  and explore the performance impacts when each worker runs OpenMC on 1, 2, 4, 8, and 16 nodes with 8 GPUs per node. Because Crusher has a total of 192 nodes, if we choose more than 4 workers (e.g., 8, 10, or 16), there are not enough available compute nodes and number of different experiments for using ytopt-libe. 

\begin{figure}[ht]
\center
  \includegraphics[width=\linewidth]{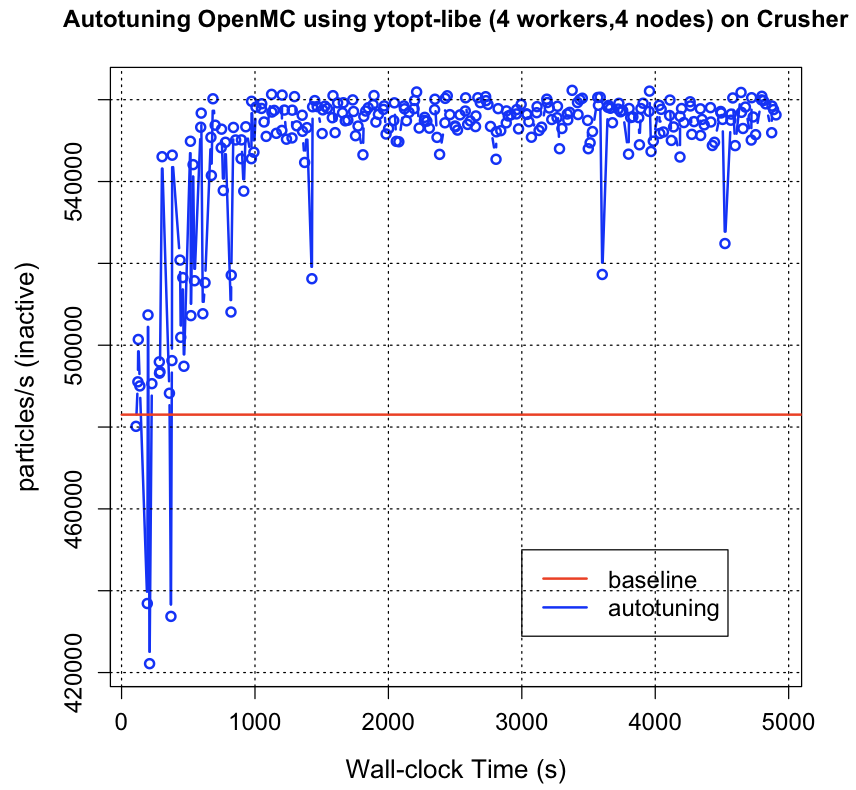}
 \caption{Autotuning OpenMC on 8 GPUs using ytopt-libe}
\label{fig:yl8}
\end{figure}

Figure \ref{fig:yl8} shows autotuning OpenMC on 8 GPUs using ytopt-libe with 4 workers on 4 nodes with 8 GPUs per node. Over time, ytopt-libe leads to the high-performing regions of the parameter space, which results in much better performance than the baseline performance and identifies the best configuration, which results in 16.41\% performance improvement. 

\begin{figure}[ht]
\center
  \includegraphics[width=\linewidth]{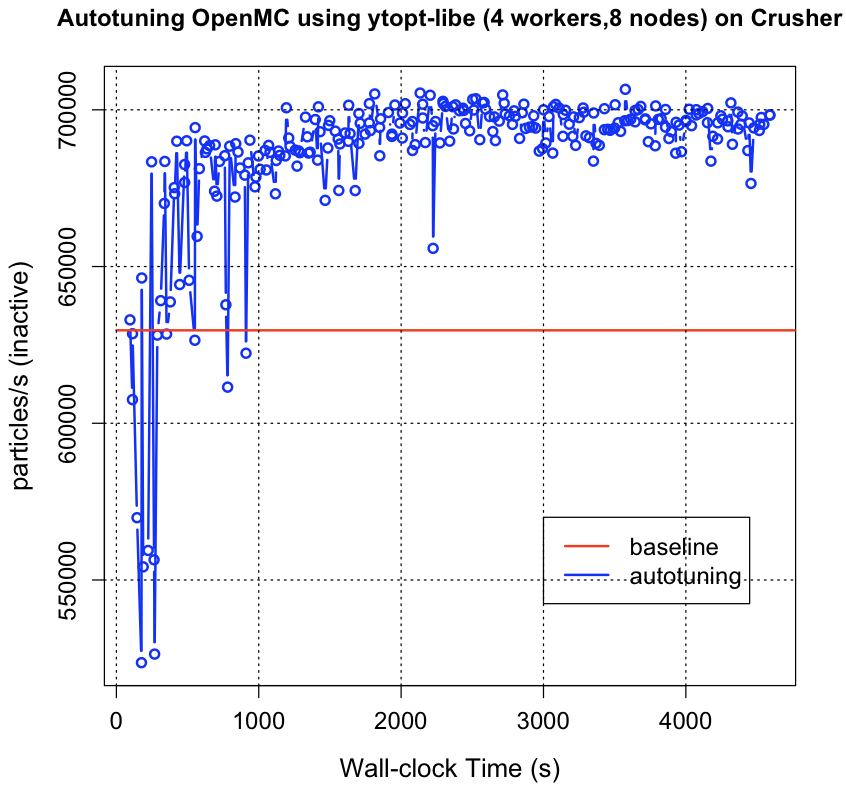}
 \caption{Autotuning OpenMC on 16 GPUs using ytopt-libe}
\label{fig:yl16}
\end{figure}

\begin{figure}[ht]
\center
  \includegraphics[width=\linewidth]{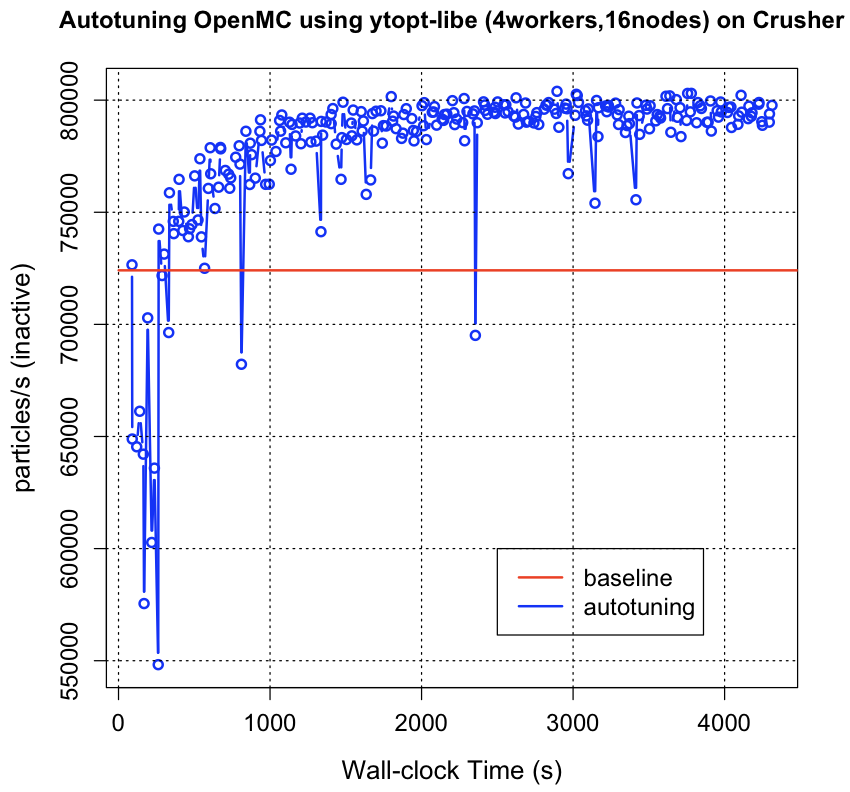}
 \caption{Autotuning OpenMC on 32 GPUs using ytopt-libe}
\label{fig:yl32}
\end{figure}

\begin{figure}[ht]
\center
  \includegraphics[width=\linewidth]{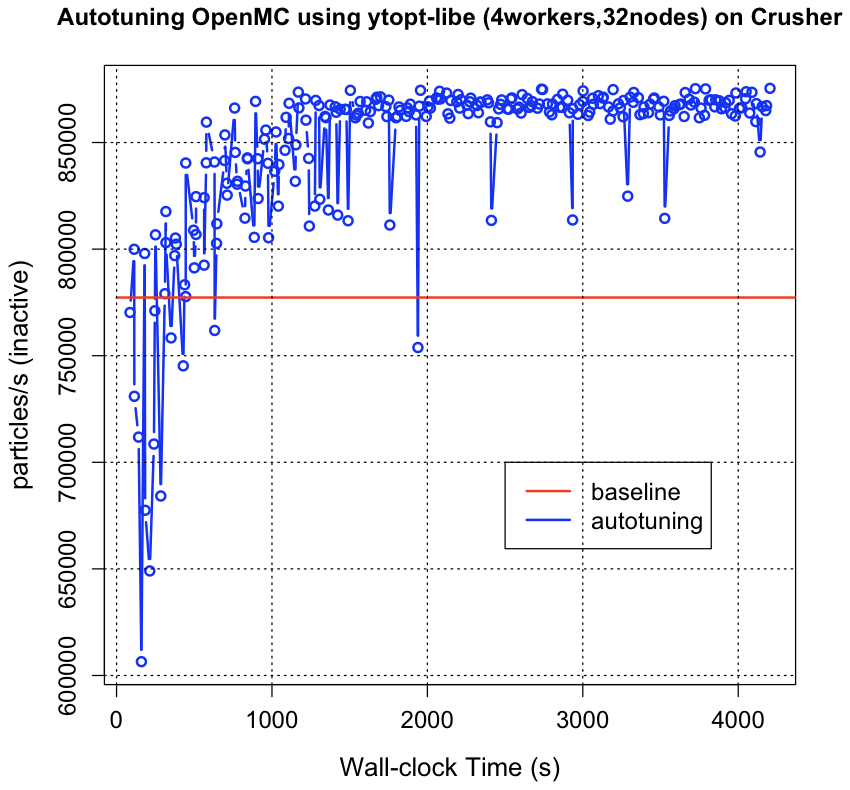}
 \caption{Autotuning OpenMC on 64 GPUs using ytopt-libe}
\label{fig:yl64}
\end{figure}

\begin{figure}[ht]
\center
  \includegraphics[width=\linewidth]{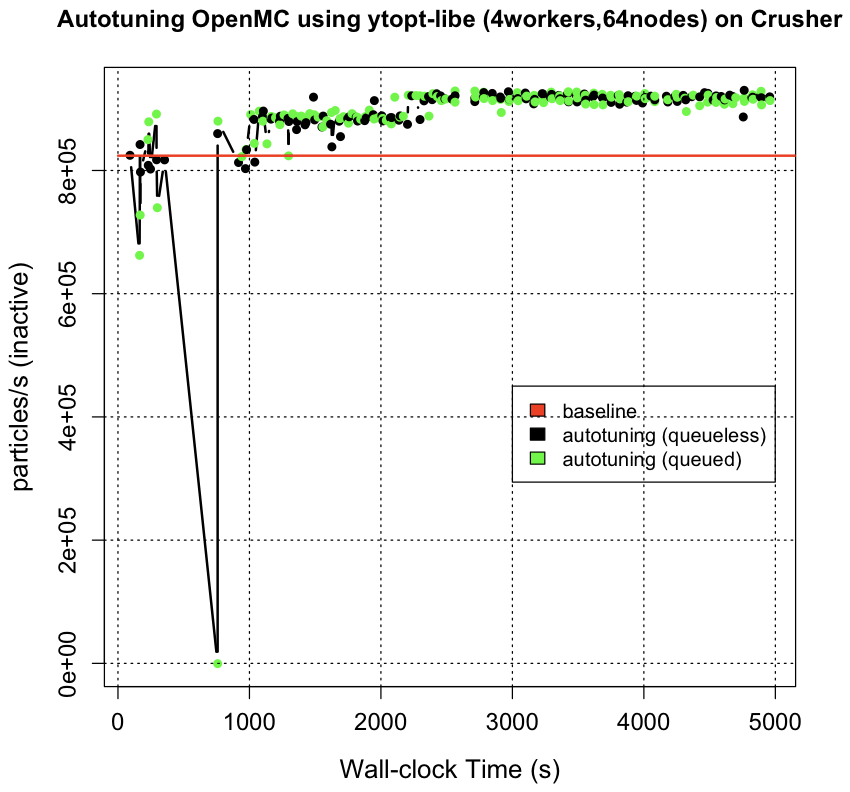}
 \caption{Autotuning OpenMC on 128 GPUs with queued and queueless using ytopt-libe}
\label{fig:yl128}
\end{figure}

Similarly, to autotune OpenMC on different numbers of GPUs with  256 evaluations, we conduct the following experiments. We configure ytopt-libe  to run with 4 workers on 8 nodes (Figure \ref{fig:yl16}), on 16 nodes  (Figure \ref{fig:yl32}), on 32 nodes (Figure \ref{fig:yl64}), and on 64 nodes (Figure \ref{fig:yl128}). Table \ref{tab:ylpc} summarizes the performance improvement comparison using topt-libe with 4 workers and different numbers of GPUs on Crusher. We observe that  ytopt-libe identifies the best configuration, which results in at least 11.03\% performance improvement. However, the scalablity of OpenMC from 8 GPUs to 128 GPUs is limited because the design for the version of OpenMC mainly focuses on the single-GPU performance.

\begin{table*}[ht]
\small\sf\centering
\caption{Performance  improvement for OpenMC on different numbers of GPUs using ytopt-libe with 4 workers}
\begin{tabular}{llll}
\toprule
Number of GPUs & Baseline Performance (Particles/s) & Best Performance (Particles/s)& Performance improvement (\%)  \\
\midrule
8 &  483033  & 562288  & 16.41 \\
16  & 629647 & 706535  &  12.21\\
32  & 724098 & 803931  & 11.03\\
64  & 777287 & 875419  & 12.62\\
128  & 823997 & 930078  & 12.87\\
\bottomrule
\end{tabular}
\label{tab:ylpc}
\end{table*}

Figure \ref{fig:yl128} shows autotuning OpenMC on 8 GPUs using ytopt-libe with 4 workers on 64 nodes with 8 GPUs per node. Over time, ytopt-libe leads to the high-performing regions of the parameter space, which results in much better performance than the baseline performance and identifies the best configuration, which results in 12.87\% performance improvement. However, OpenMC with one configuration had a timeout termination. so  that one negative value was assigned by ytopt-libe as the worst parameter space region. In this figure we split the overall autotuning process into two parts: one with the color green for OpenMC with the queued mode and the other with the color black for OpenMC with the queueless mode. We observe that in the high-performing regions of the parameter space, using OpenMC with the queued or queueless mode does not impact the performance much. This indicates that queued and queueless modes are not main performance factor for OpenMC.

\section{Autotuning in Runtime, Energy, and EDP}

In the preceding section the results are all for the application FoM (particles/s) as the performance metric. What about other performance metrics? Application runtime (without the initialization time) is used to calculate the rate particles/s for the FoM. Energy consumption is the product of power and runtime and it captures the trade-off between power consumption and runtime. The lowest energy means the best tradeoff between runtime and power. The energy delay product is the product of energy and runtime or the product of power and square of runtime, and it captures the trade-off between energy and runtime. The lowest EDP means the best tradeoff between energy and runtime. In this section we apply the proposed autotuning framework ytopt-libe in Figure~\ref{fig:yl} to autotune OpenMC in the application runtime, energy, and EDP on Crusher. We use the autotuning framework to explore the tradeoffs for the application's efficient execution with these performance metrics.

\xingfu{For measuring the baseline energy for each application executed on 8 GPUs on a single node, we use APEX \cite{R8} to run the application with the default application settings from Table \ref{tab:tp} on the default system environment (1 MPI rank per GPU, 8 threads per CPU, omp placement: threads) five times. Then we use the smallest runtime/energy as the baseline runtime/energy.} 

\begin{figure}[ht]
\center
 \includegraphics[width=\linewidth]{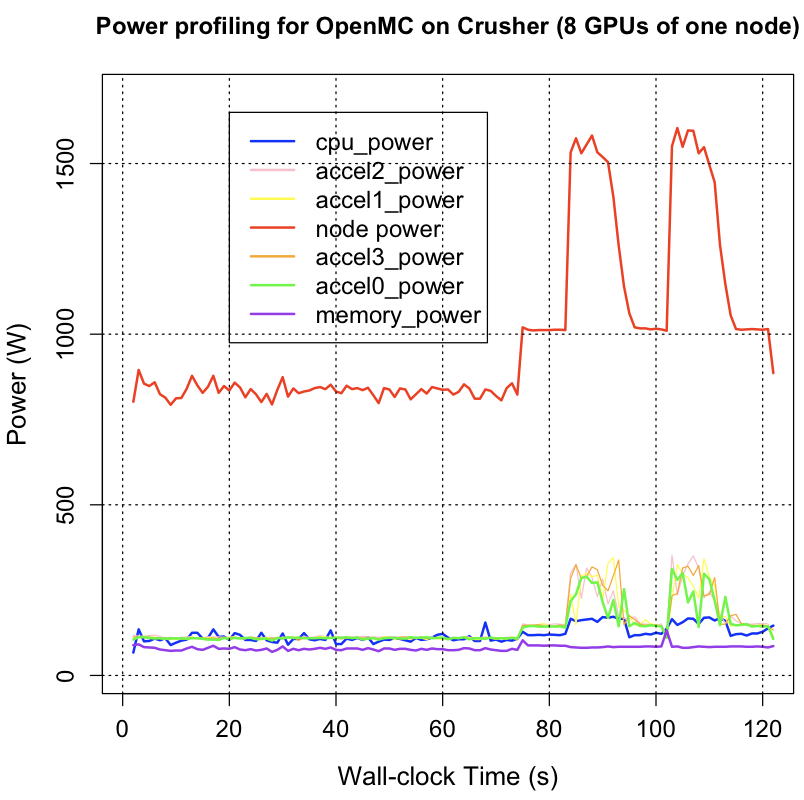}
 \caption{Power profiling for running OpenMC on 8 GPUs of a single node on Crusher}
\label{fig:p8}
\end{figure}

Figure \ref{fig:p8} shows the power profiling for OpenMC with the default application settings on 8 GPUs of a single node using APEX. Comparing this figure with Figure \ref{fig:p1}, we can see that the default settings are not sufficient for efficient execution of OpenMC on 8 GPUs because the number of particles in-flight is not large enough to keep the GPUs busy for the first and second iterations. Loading the HD5 datasets takes more than 70 s at the application initialization and is the performance bottleneck as well.

\begin{figure}[ht]
\center
  \includegraphics[width=\linewidth]{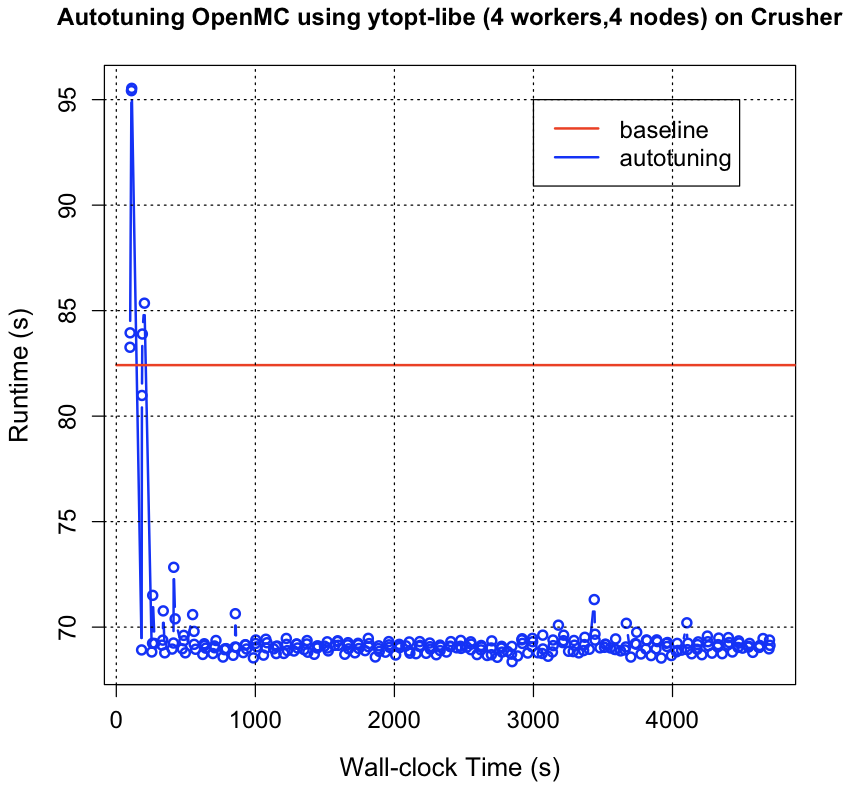}
 \caption{Autotuning OpenMC with the metric runtime on 8 GPUs using ytopt-libe}
\label{fig:ylr}
\end{figure}

Figure \ref{fig:yl8} shows autotuning OpenMC in FoM on 8 GPUs using ytopt-libe with 4 workers on 4 nodes with 8 GPUs per node. For the sake of simplicity, we  use the same settings to autotune OpenMC in runtime, energy, and EDP on 8 GPUs using ytopt-libe with 4 workers on 4 nodes. Figure \ref{fig:ylr} presents autotuning OpenMC with the metric runtime on 8 GPUs using ytopt-libe. We observe that over time ytopt-libe leads to the high-performing regions of the parameter space, which results in  much better runtime than that of the baseline and identifies the best configuration, which results in 17.05\% performance improvement. 

For autotuning energy and EDP, after the evaluation of a configuration APEX generates a summary report that records the package energy and DRAM energy for each node. We accumulate these values as the node energy. When ytopt receives the report from APEX, it 
calculates an average node energy and EDP and uses the average energy or EDP as the primary metric for autotuning. 

\begin{figure}[ht]
\center
  \includegraphics[width=\linewidth]{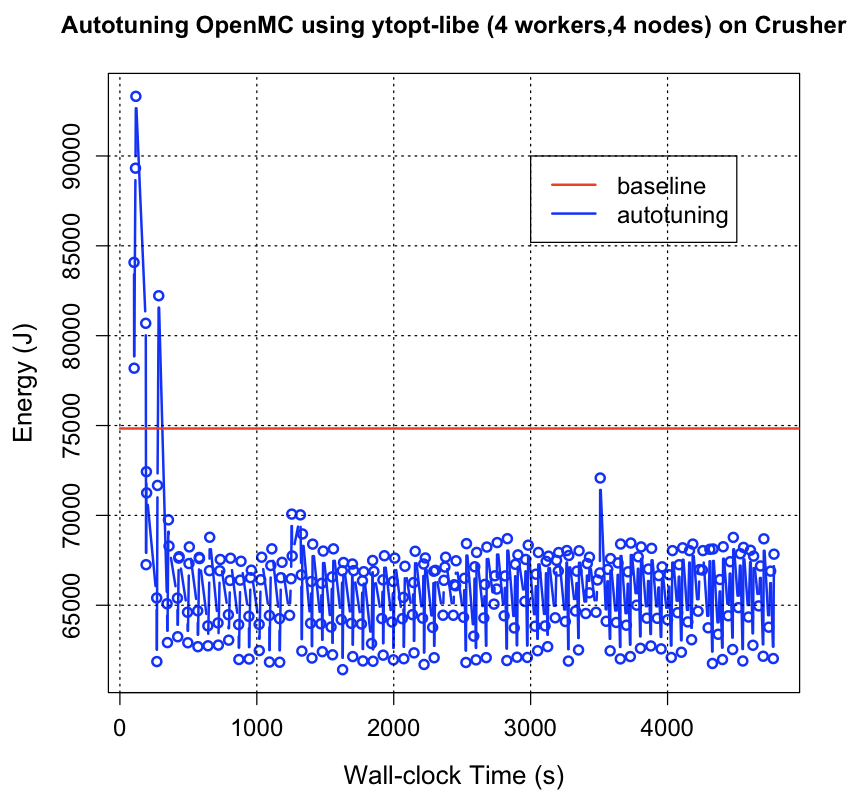}
 \caption{Autotuning OpenMC with the metric energy on 8 GPUs using ytopt-libe}
\label{fig:yle}
\end{figure}

Figure \ref{fig:yle} shows autotuning OpenMC with the metric energy on 8 GPUs using ytopt-libe. Although ytopt-libe identifies the best configuration, which results in 17.93\% energy improvement, we observe that the energy pattern over time is  not approximate to one good point. This is very different from Figure \ref{fig:ylr} where the runtime pattern is approximate to the best point with very little margin. This indicates that when ytopt-libe used four different nodes to run 4 evaluations for OpenMC in parallel, the power measurement for each node was varied significantly so that the energy consumption for each node had a larger difference. %

\begin{figure}[ht]
\center
  \includegraphics[width=\linewidth]{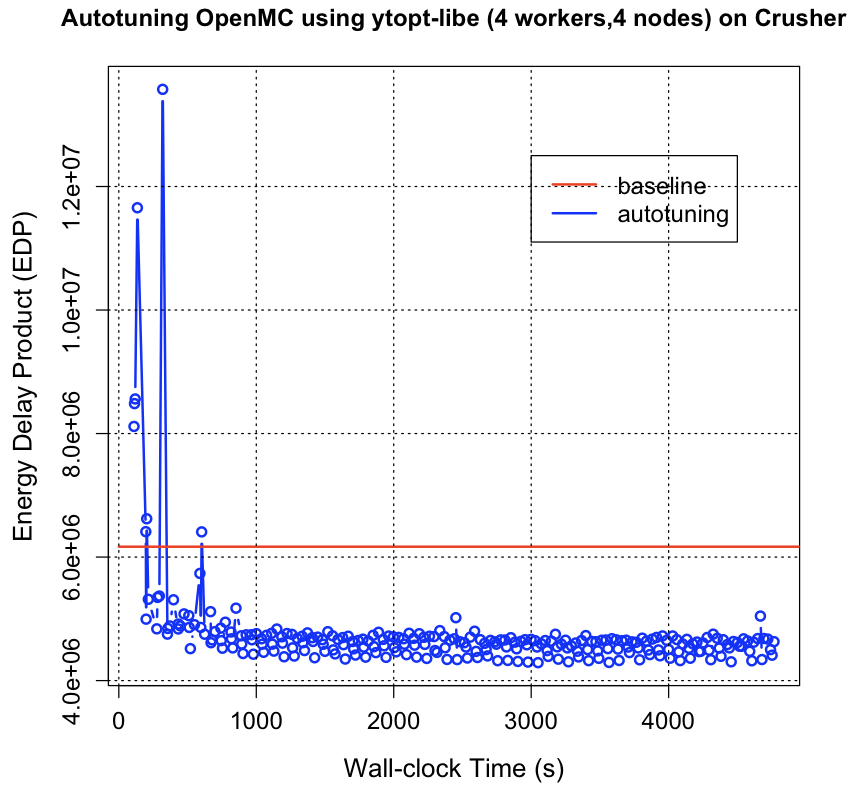}
 \caption{Autotuning OpenMC with the metric EDP on 8 GPUs using ytopt-libe}
\label{fig:yld}
\end{figure}

Further, when we use ytopt-libe to autotune OpenMC with the metric EDP in Figure \ref{fig:yld}, ytopt-libe identifies the best configuration, which results in 30.44\% EDP improvement. The reason for this significant improvement is that EDP is the product of energy and runtime and the product of power and square of runtime, and the square of runtime makes the EDP trend stable.

\section{Conclusions}

In this paper we \xingfu{extended and enhanced our ytopt autotuning framework to} present our methodology and framework ytopt-libe to integrate ytopt and libEnsemble to \xingfu{tackle ytopt's limitations and} take advantage of massively parallel resources to accelerate the autotuning process. Specifically, we focused on using the proposed framework to autotune the application OpenMC. OpenMC has seven tunable parameters,  some of which have large ranges, and it is time-consuming to set the proper combination of these parameter values to achieve the best performance. We applied the proposed framework to autotune the MPI/OpenMP offload version of OpenMC based on a user-defined metric such as the FoM  (particles/s) or runtime, energy, and EDP on the OLCF Frontier TDS system Crusher. The experimental results show that we achieved an improvement of up to 29.49\% in FoM and up to 30.44\% in EDP. 

\begin{figure}[ht]
\center
  \includegraphics[width=\linewidth]{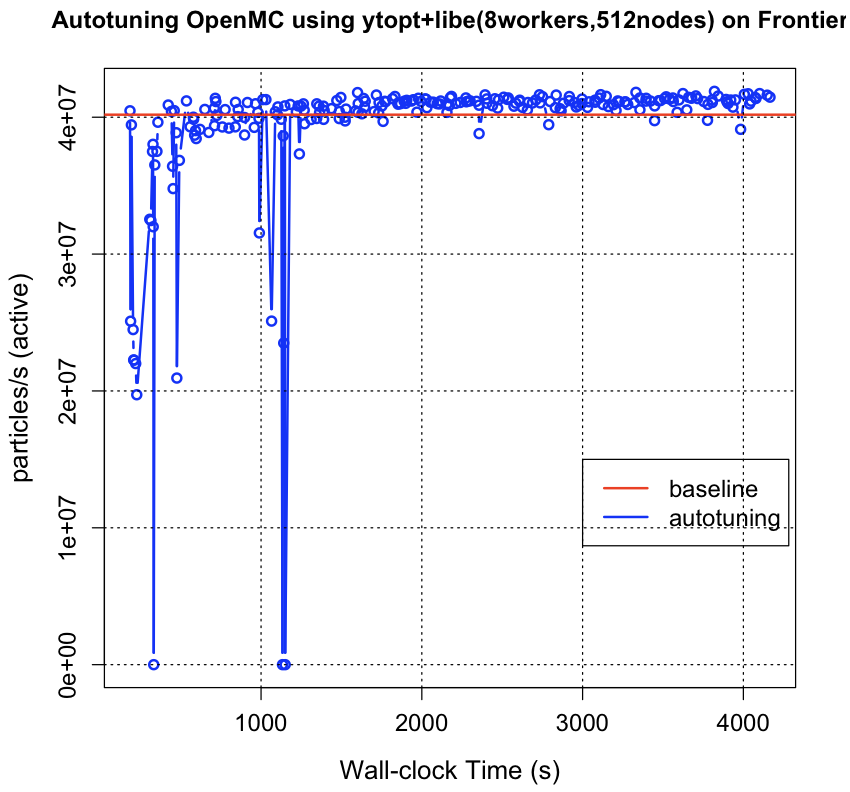}
 \caption{Autotuning OpenMC on 512 GPUs (64 nodes) using ytopt-libe with 8 workers}
\label{fig:of}
\end{figure}

\xingfu{In our previous work \cite{R1} we used ytopt to autotune four hybrid MPI/OpenMP ECP proxy applications \cite{R21} at large scales on large-scale HPC systems Cray XC40 Theta  at Argonne National Laboratory and the IBM Power9 heterogeneous system Summit at Oak Ridge National Laboratory. Recently, we applied ytopt-libe to autotune the latest development version (2023.09.08) of OpenMC (which removes queue and queueless modes) on 512 GPUs (64 nodes) per worker with the total 512 nodes for 8 workers  on the exascale system OLCF Frontier \cite{R51} in Figure \ref{fig:of}, ytopt-libe identified the best configuration to result in the best performance. ytopt-libe is a ML-based Python software package and is available from our GitHub repo https://github.com/ytopt-team/ytopt-libensemble, and the autotuning scripts for OpenMC used in this paper are available from our GitHub repo https://github.com/ytopt-team/ytopt-libensemble/tree/main/ytopt-libe-openmc. Our autotuning framework is not dependent on applications with any kinds of languages and any types of systems. As long as a user defines a parameter space with tunable application parameters and system parameters, ytopt-libe is able to identify the best configuration to result in the best performance. This autotuning is done only once. Then the application with this best configuration can be used for efficient application execution at large scales with the smallest runtime and/or the lowest energy.}

\section{Acknowledgements}
This work was supported in part by the DOE ECP PROTEAS-TUNE project, in part by the DOE ASCR SciDAC RAPIDS2 and OASIS, in part by the ECP PETSc/TAO project, and in part by the ECP ExaSMR Project.  We acknowledge the Oak Ridge Leadership Computing Facility for use of Crusher and Frontier under the projects CSC383 and Kevin Huck from University of Oregon for the power measurement support of APEX on Crusher. This material is based upon work supported by the U.S. Department of Energy, Office of Science, under contract number DE-AC02-06CH11357, at Argonne National Laboratory.

\end{document}